\documentclass[%
 aip,
 amsmath,amssymb,
 reprint,%
]{revtex4-1}

\usepackage{graphicx}% Include figure files
\usepackage{dcolumn}% Align table columns on decimal point
\usepackage{bm}% bold math
\usepackage[utf8]{inputenc}
\usepackage[T1]{fontenc}
\usepackage{mathptmx}
\usepackage{etoolbox}
\usepackage{xcolor}
\usepackage{amssymb,amsfonts,amsmath}
\usepackage[english]{babel}
\usepackage[colorlinks=true,linkcolor=blue,citecolor=blue]{hyperref}%
\makeatletter
\def\@email#1#2{%
 \endgroup
 \patchcmd{\titleblock@produce}
  {\frontmatter@RRAPformat}
  {\frontmatter@RRAPformat{\produce@RRAP{*#1\href{mailto:#2}{#2}}}\frontmatter@RRAPformat}
  {}{}
}%
\makeatother
\begin{document}
\title{Flocking and swarming in a multi-agent dynamical system}
	
\author{Gourab Kumar Sar}
\email{mr.gksar@gmail.com}
\affiliation{Physics and Applied Mathematics Unit, Indian Statistical Institute, 203 B. T. Road, Kolkata 700108, India}
\author{Dibakar Ghosh}
\email{dibakar@isical.ac.in}
\affiliation{Physics and Applied Mathematics Unit, Indian Statistical Institute, 203 B. T. Road, Kolkata 700108, India}	

%\subject{epidemic spreading, complex network}

%\keywords{Complex network, epidemic spreading, final outbreak size, test-kit}
%\thanks{These two authors contributed equally}
	
\thanks{Corresponding Auhtor: Gourab Kumar Sar}
%\email{mr.gksar@gmail.com}

%%%% Abstract text to be placed here %%%%%%%%%%%%
\begin{abstract}
\par {Over the past few decades, the research community has been interested in the study of multi-agent systems and their emerging collective dynamics. These systems are all around us in nature, like bacterial colonies, fish schools, bird flocks, as well as in technology, such as microswimmers and robotics, to name a few. Flocking and swarming are two key components of the collective behaviours of multi-agent systems. In flocking, the agents coordinate their direction of motion, but in swarming, they congregate in space to organise their spatial position. We investigate a minimal mathematical model of locally interacting multi-agent system where the agents simultaneously swarm in space and exhibit flocking behaviour. Various cluster structures are found, depending on the interaction range. When the coupling strength value exceeds a crucial threshold, flocking behaviour is observed. We do in-depth simulations and report the findings by changing the other parameters and with the incorporation of noise.}
\end{abstract}

% \pacs {05.45.Xt, 05.45.Gg, 85.25.Cp, 87.19.lm}
\maketitle
\begin{quotation}
{\color{blue} This article is in honour of 70th birthday of Prof. Juergen Kurths. The objective of this work is an extension of his works on mobile oscillators.}
\\
\\
Multi-agent systems are ubiquitously found in nature in the form of school of fish, honey bees, locust swarms etc. The study of their self-organization and movement remains of huge interest to researchers. What is fascinating is the emergence of coordinated movements without any central controller. In this paper, we propose a minimal model which captures coordination of the agents through local neighbor interactions. We are interested in exploring both the spatial formations as well as the synchronization of moving directions. Many past studies on such systems have either focused on the spatial dynamics ignoring the directions of movements or reported the directional synchrony without shedding much light on the positions of the agents. We have tried to reduce this gap by investigating the simultaneous occurrence of swarming and flocking, i.e., by focusing both on the agents' positions and directions. These agents are also free to move in the unbounded plane without any boundaries. This relaxation makes our model more realistic as the movements of animals or birds are often unrestricted.

\end{quotation}
%\begin{fmtext}

%%%%%%%%%%%%%%%%%%%%%%%%%%%%%%%%%%%%%%%%%%%%%%%%%%%%%%
\section{Introduction}
%{\centering  }
%\end{fmtext} 
%\maketitle
Why do animals like sheep and deer move in herds and starlings flock together as they fly? There are some obvious reasons underlying this behavior. Living entities characteristically live in colonies by maintaining connections with nearby ones. This helps them when migrating from one place to another, increases the opportunity to locate food, protects them from predator attacks etc. The more important question though is how they do it. In this pursuit, researchers started studying the motion of flocks of birds, herd of sheep, fish schools, ant colonies etc.~\cite{bialek2012statistical,hemelrijk2012schools,sumpter2010collective} These ostensibly intriguing behaviours may be reduced to two primary categories. The trivial one is the spatial mobility and other is the internal feedback among them. Over the last few decades biologists, physicists, computer scientists, and control theorists have tried to comprehend the motion and cooperation of these entities~\cite{reynolds1987flocks,toner1998flocks,okubo1986dynamical,norris1988cooperative,cucker2007emergent,toner2005hydrodynamics,nagy2010hierarchical}. It is believed that the understanding of the natural instinctive collaboration among these agents might help us taking better social decisions, and in communications~\cite{miller2010smart,al2018bat,shklarsh2011smart}. Research fields like \textit{smart swarm}~\cite{miller2010smart}, \textit{multi-agent systems}~\cite{van2008multi}, \textit{swarm robotics}~\cite{brambilla2013swarm}, \textit{swarm intelligence}~\cite{kennedy2006swarm} etc. have often dedicated a special attention towards these natural behaviors. Overall, it has piqued the intense curiosity of researchers from diverse fields.

\par One of the pioneering works in this area was carried out by Craig Reynolds in 1987~\cite{reynolds1987flocks}. Reynolds introduced the concept of "boids", which are simple simulated entities that follow three basic rules: separation (avoidance of collisions with nearby entities), alignment (matching the velocity and direction of nearby entities), and cohesion (moving toward the average position of nearby entities). Reynolds' findings set the groundwork for additional study and served as an inspiration for numerous additional investigations into the social behaviour of animals. More complex models have been created over time to replicate and comprehend the dynamics of flocks and swarms. These models frequently include extra elements like effects from the environment, personal decision-making processes, and inter-element communication~\cite{vicsek2012collective}. Computer scientists use a set of rules called algorithms to simulate the behavior of such systems, whereas, physicists, biologists, and mathematicians take the path of dynamical modelling to describe and analyze the behavior of systems over time. Algorithms are step-by-step procedures or rules used to solve specific problems or simulate individual behavior~\cite{hartman2006autonomous}. The goal of dynamic modelling, on the other hand, is to explain and anticipate emergent behaviour by creating mathematical models that describe the dynamic interactions of a system across time~\cite{carrillo2010particle}. Instead of laying down specific guidelines for each behaviour, it concentrates on the dynamics of the entire system.

\par In 1995, Vicsek et al.~\cite{vicsek1995novel} proposed such a dynamical model and studied the collective motion in self-propelled particle systems. It assumes particles moving in a continuous space, interacting with nearby particles to align their directions of motion. They showed the transition from a gas like incoherent state where the particles' directions are disordered, to a ordered state in which the particles move with coherent directions. Models related to circadian rhythms of plants and animals were introduced much before the Vicsek model by Winfree~\cite{winfree1967biological} which was later followed by the pioneering work of Kuramoto~\cite{kuramoto1975self}. Their works gave rise to the field \textit{synchronization}~\cite{pikovsky2001universal}. The ubiquitous occurrence of this synchronization phenomena in various natural~\cite{buck1988synchronous,neda2000sound,womelsdorf2007role,schafer1998heartbeat} and man-made~\cite{rohden2012self,sichitiu2003simple,little1966synchronization} systems meant that it caught the rightful attention of scientists. However, most of the studies on synchronization phenomena do not care about the oscillator's spatial position. On a parallel front, works involving the spatial movements like aggregation structures and swarming dynamics were also carried out~\cite{bernoff2013nonlocal,topaz2006nonlocal,topaz2004swarming}.
The effect of agents' mobility on synchronization of internal dynamics has been investigated while studying the dynamics of \textit{moving agents}~\cite{frasca2008synchronization,chowdhury2019synchronization,majhi2019emergence}. Recent research has looked at the interactions between internal phases and spatial positions, and these entities are known as \textit{swarmalators}~\cite{o2017oscillators,sar2022dynamics,sar2023pinning,sar2022swarmalators,sar2023solvable}. Over the years, researchers have used both deterministic~\cite{cucker2007emergent,chen2014minimal,kolokolnikov2013emergent,liu2009synchronization} and stochastic~\cite{reynolds2022stochastic,reynolds2023stochasticity,reynolds2023swarm,ahn2010stochastic,morin2015collective} modelling approaches while investigating these features. Reynolds et al.~\cite{reynolds2022stochastic} introduced a stochastic model where they studied spatial cohesiveness and collective motion along with reporting empirical evidences in jackdaw flocks.
Studies have also been carried out by conducting laboratory and field experiments to delve deeper into this topic~\cite{kelley2013emergent,aihara2008mathematical,aihara2014spatio}.

\par Most of the works related to the locomotion of self-propelled agents have been performed by considering the agents lying inside a confined region. This consideration keeps the agents bounded and enables computational and analytical explorations in a controlled environment. In this articles, we consider a multi-agent dynamical system where the agents are free to move in the two dimensional plane without any bound. Moreover, in place of the simple spatial dynamics controlled by the velocity, we have considered spatial interactions among the agents which captures their swarming behavior. Each agent interacts locally with all others lying inside a circular interaction range determined by the sensing radius. The interaction with the local neighbors takes place both in the spatial and directional dynamics. We study the impact of the sensing radius on the spatial formation. We show that depending on its value the agents can form disjoint clusters or move in a single connected cluster. We investigate the effect of directional coupling strength on flocking of the agents. The effect of surrounding noise has also been taken into consideration.

%%%%%%%%%%%%%%%%%%%%%%%%%%%%%%%%%%%%%%%%%%%%%%%%%%%%%
\section{\label{sec:level2}Two collective behaviors: Swarming and flocking}
\begin{figure}[t]
	\centerline{
		\includegraphics[width = \columnwidth]{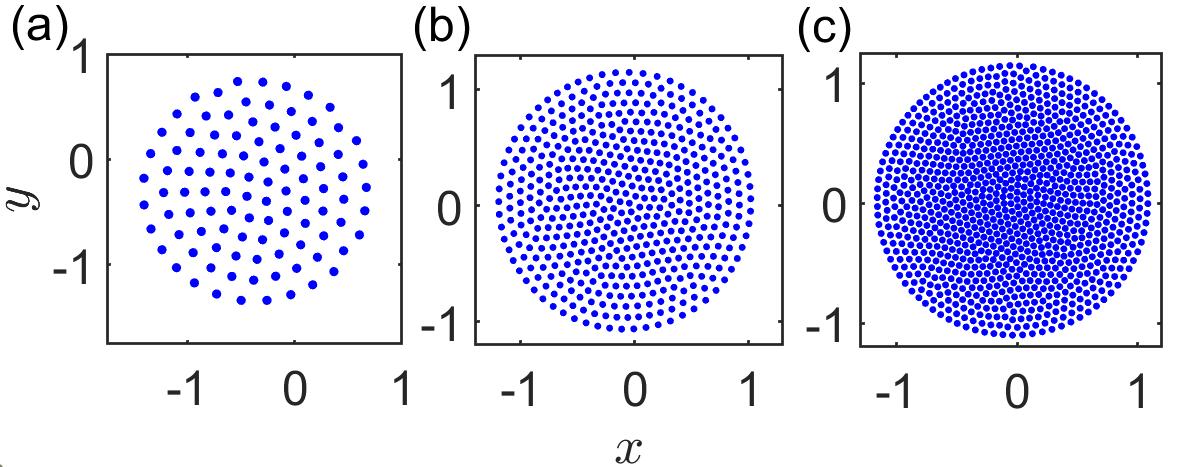}}
	\caption{Aggregation structure governed by the dynamics given by Eq.~(\ref{eq.2}) with $A=B=1$. The particles are initially chosen from $[-1,1] \times [-1,1]$ randomly. Simulations are performed for (a) $N =100$, (b) $N = 500$, and (c) $N = 1000$ particles.}
	\label{fig1}
\end{figure}

First, we briefly discuss these two collective phenomena, namely, \textit{swarming} and \textit{flocking} which are useful in the context of our work. As mentioned earlier, both these behaviors have much in common. Both evolve as a result of interactions among large number of agents/organisms. Now, consider that each of these agents has specific spatial position $\textbf{x}_{i} \in \mathbb{R}^{d}$ where they can be located, and a direction $\theta_i$ for movement. If we focus mainly on the spatial movements of the agents neglecting their directions of movement, it is considered as \textit{aggregation} which is the key feature of swarming. We mention one such aggregation model below,
\begin{equation}
	\Dot{\textbf{x}}_{i} =\frac{1}{N} \sum_{\substack{j = 1\\j \neq i}}^{N}\left[ F_{att}(\textbf{x}_{j}-\textbf{x}_{i}) - F_{rep}(\textbf{x}_{j}-\textbf{x}_{i}) \right],
	\label{ eq.1}
\end{equation}
for $i=1,2,...,N$. $F_{att}$ and $F_{rep}$ are the spatial attraction and repulsion forces between the agents, respectively, and $N$ is the total number of agents. The attraction force keeps the agents close to each other by making sure that they do not disperse indefinitely. On the other hand, the repulsive force ensures that the agents do not collide or be asymptotically close. A continuum limit study of this individual-based particle model was carried out by Fetecau et al.~\cite{fetecau2011swarm} in terms of an integro-differential equation. If we specifically choose linear attraction and power-law repulsion kernels, then the aggregation model becomes
\begin{equation}
	\dot{\textbf{x}}_{i} =\frac{1}{N} \sum_{\substack{j = 1\\j \neq i}}^{N}\left[ A (\textbf{x}_{j}-\textbf{x}_{i}) - B  \frac{\textbf{x}_{j}-\textbf{x}_{i}}{{|\textbf{x}_{j}-\textbf{x}_{i}|}^{2}} \right].
	\label{eq.2}
\end{equation} 
We simulate this model in two-dimension with $A=B=1$ and demonstrate the aggregation structure in Fig.~\ref{fig1}. It is found that the particles arrange themselves in a disc-like structure maintaining a minimal distance from each other. The radius of the disc in the thermodynamic limit $N \to \infty$ is found to be $\sqrt{B/A}$~\cite{o2017oscillators}.

Now, moving to the direction of movement of the agents $\theta_i$, one can measure their tendency to move in a flock or align their orientation in space. This is referred to as \textit{alignment} and is the other component of swarming. Consider the following model,
\begin{align}
	\dot{\textbf{x}}_{i} &=v\left[
	\begin{array}{c}
		\cos \theta_i\\
		\sin \theta_i
	\end{array}\right],
	\label{eq.3}\\
	\dot{\theta}_{i} &= u_{i}
	\label{eq.4},
\end{align}
where $\textbf{x}_{i} = (x_i,y_i) \in \mathbb{R}^{2}$, $v$ is the self-propulsion speed, and $u_i$ is some external control like the angular velocity of the agent. Alignment is basically synchronization of agents' directions and this is regarded as \textit{flocking} in literature. The flocking problem associated with the model Eqs.~(\ref{eq.3})-(\ref{eq.4}) is defined as follows~\cite{zhu2012flocking,liu2008connectivity,olfati2006flocking}: \textit{flocking} is said to be achieved if all the agents' direction become same at large time and collision among them is always avoided, i.e., 
\begin{equation}
	|\textbf{x}_{j}-\textbf{x}_{i}| > 0 \hspace{0.5cm} \forall t \in [0, +\infty],
	\label{eq.5}
\end{equation}
\centerline{and}
\begin{equation}
	\lim_{t \to +\infty} (\theta_{i}(t)-\theta_{j}(t)) = 0,
	\label{eq.6}
\end{equation}
for all $1 \leq i,j \leq N$. From the discussion so far, it is perceived that swarming is mostly about aggregation of spatial position, whereas flocking deals with direction synchronization. A desire to stick with the group and a drive to prevent collisions appear to be two balanced and opposite tendencies that make up natural flocks and swarms. The first is a necessity for some social activities, such as protection from predators, migration, and having more opportunities to locate food. The latter is to maintain enough room for themselves, for instance, the birds need some space for flapping wings.

%%%%%%%%%%%%%%%%%%%%%%%%%%%%%%%%%%%%%%%%%%%%%%%%%%%%
\section{\label{sec:level3}Model Description}
In this paper we study flocking and swarming together. We investigate how the agents move in space while avoiding collision at all times, how their directions change, and the circumstances under which they achieve flocking while forming various spatial structures. The agents move in infinite plane with a heading direction. At an instant of time, it interacts with its neighboring agents by both position and direction components, i.e., an agent influences the position and direction of other agents which are its neighbor. At time $t$, by neighbors of agent $i$ we mean those which lie inside a circle of radius $r > 0$ centred at $\textbf{x}_{i}(t)$. This radius is often called \textit{sensing/interaction/vision radius}. The neighbor set of agent $i$ is described by
\begin{equation}
	N_i = \{j \neq i \hspace{0.2cm}|\hspace{0.2cm} |\textbf{x}_{j}-\textbf{x}_{i}|<r \}.
	\label{eq.7}
\end{equation}
The agents are not neighbors of themselves and the ones which lie outside the sensing radius. We can characterize the neighbor relationship of the agents by an undirected graph $G = \{V, E, A\}$, where the vertex/node set $V = \{1,2,\ldots,N\}$ are the indices of the agents, the edge/link set is given by 
\begin{equation}
	E = \{(i,j) \subset V \times V \hspace{0.2cm} | \hspace{0.2cm} |\textbf{x}_{j}-\textbf{x}_{i}|<r,\hspace{0.2cm} i \neq j\}. \label{eq.8}
\end{equation}
The adjacency matrix $A = (a_{ij})$ is an $N\times N$ matrix defined as
\begin{equation}
	a_{ii} = 0,\hspace{0.2cm} a_{ij} =
	\begin{cases}
		1, & \text{$|\textbf{x}_{j}-\textbf{x}_{i}|<r$},\\
		0, & \text{$|\textbf{x}_{j}-\textbf{x}_{i}|\geq r$},
	\end{cases} \hspace{0.2cm} (i \neq j) \label{eq.9}
\end{equation}
The graph $G$ is called proximity graph. The Laplacian matrix of the graph $G$ is defined by another $N \times N$ matrix $\mathcal{L} = (l_{ij})$, where
\begin{equation}
	l_{ii} = \sum_{\substack{j = 1,j \neq i}}^{N} a_{ij}, \hspace{0.2cm} l_{ij} = -a_{ij} \hspace{0.2cm} (i \neq j). \label{eq.10}
\end{equation}
For an undirected graph $G$, its Laplacian matrix $\mathcal{L}$ is symmetric and positive semi-definite. $\mathcal{L}$ has at least a zero eigenvalue with eigenvector $\mathbf{1}_N$, i.e., $\mathcal{L}\mathbf{1}_N = 0$. The eigenvalues of $\mathcal{L}$ are also called the eigenvalues of graph $G$ and are described by
\begin{equation}
	0 = \lambda_1(G)\leq \lambda_2(G) \leq \ldots \leq \lambda_n(G).\label{eq.11}
\end{equation}
It is well known that $G$ is connected if and only if $\lambda_2(G)$ is positive. $\lambda_2(G)$ is usually called the \textit{algebraic connectivity}. Large value of $\lambda_2(G)$ implies a strong connection of the graph $G$. Finally, we define our mathematical model as
\begin{align}
	\dot{\mathbf{x}}_{i} &=v\left[
	\begin{array}{c}
		\cos \theta_i\\
		\sin \theta_i
	\end{array}\right]+
	\frac{1}{|N_{i}|} \sum_{\substack{j = 1\\j \neq i}}^{N}a_{ij}\left[ \frac{\mathbf{x}_{j}-\textbf{x}_{i}}{|\textbf{x}_{j}-\textbf{x}_{i}|}  - \frac{\textbf{x}_{j}-\textbf{x}_{i}}{{|\textbf{x}_{j}-\textbf{x}_{i}|}^{2}} \right],
	\label{eq.12}\\
	\dot{\theta}_{i} &=\frac{K}{|N_{i}|} \sum_{\substack{j = 1}}^{N} a_{ij} \sin(\theta_j - \theta_i), 	
	\label{eq.13}
\end{align}
for $i = 1,2,\ldots,N$. $K (>0)$ is the positive (attractive) coupling strength which minimizes the difference of heading angles among the agents. Eq.~(\ref{eq.13}) is inspired from the well-known Kuramoto model~\cite{kuramoto1975self}. Note that, the system governed by Eqs.~(\ref{eq.12})-(\ref{eq.13}) which we study to observe the simultaneous swarming and flocking behavior, is achieved by combining the swarming and flocking models which we discussed in Sec.~\ref{sec:level2}. (Here, we have replaced the linear attraction kernel in Eq.~(\ref{eq.2}) by the unit vector attraction kernel in Eq.~(\ref{eq.12}) which is useful for collision avoidance, as we will see later.)

\par There are three parameters which determine the final state of the solution, viz., constant velocity $v$, sensing radius $r$, and coupling strength $K$. $v$ controls the speed of movement of the agents in space. For larger $v$ the proximity graph $G$ changes very rapidly. Sensing radius $r$ determines the neighbors of an agent. Thus, at a time instant, $G$ depends on the position of the agents and sensing radius $r$. A large value of $K$ ensures that agents which stay nearby in space (less than $r$ distance away from each other) will coordinate their heading angles.

\par Initially, we place the agents inside the square region $[0,L] \times [0,L]$ and choose their directions randomly from $[0,2\pi]$. We integrate the model using Euler method with step size $dt=0.01$. Unless otherwise mentioned, the values $N=100$ and $L=10$ are used throughout the article. We also fix $v=0.1$. The choice of the speed of the agents is made such that they are far from being immobile and at the same time do not move too fast that they lose connection with their neighbors.

%%%%%%%%%%%%%%%%%%%%%%%%%%%%%%%%%%%%%%%%%%%%%%%%%%
\section{\label{sec:level4}Results}
As we mentioned earlier, inter-particle collision avoidance is a necessary condition for flocking. Our model given by Eqs.~(\ref{eq.12})-(\ref{eq.13}) obeys this condition at all times. This is a consequence of the spatial attraction and repulsion functions present in the model, i.e., the second term inside the summation in Eq.~(\ref{eq.12}) which is swarming part of the model. The opposing effects of the unit vector attraction $\Big(\frac{\mathbf{x}_{j}-\textbf{x}_{i}}{|\textbf{x}_{j}-\textbf{x}_{i}|}\Big)$ and power-law repulsion $\Big(\frac{\textbf{x}_{j}-\textbf{x}_{i}}{{|\textbf{x}_{j}-\textbf{x}_{i}|}^{2}}\Big)$ is responsible for this. When the agents come very close to each other, the repulsion force dominates the attraction force and restricts them from collision. On contrary, when they are far away, the attraction term dominates the repulsion term and as a result, they remain bounded. This is how inter-particle collision avoidance is guaranteed at all times. This distinguishes our model from the Vicsek type models where limited emphasis have been given on the spatial position of the particles.

\begin{figure}[ht]
	\centerline{
		\includegraphics[width = \columnwidth]{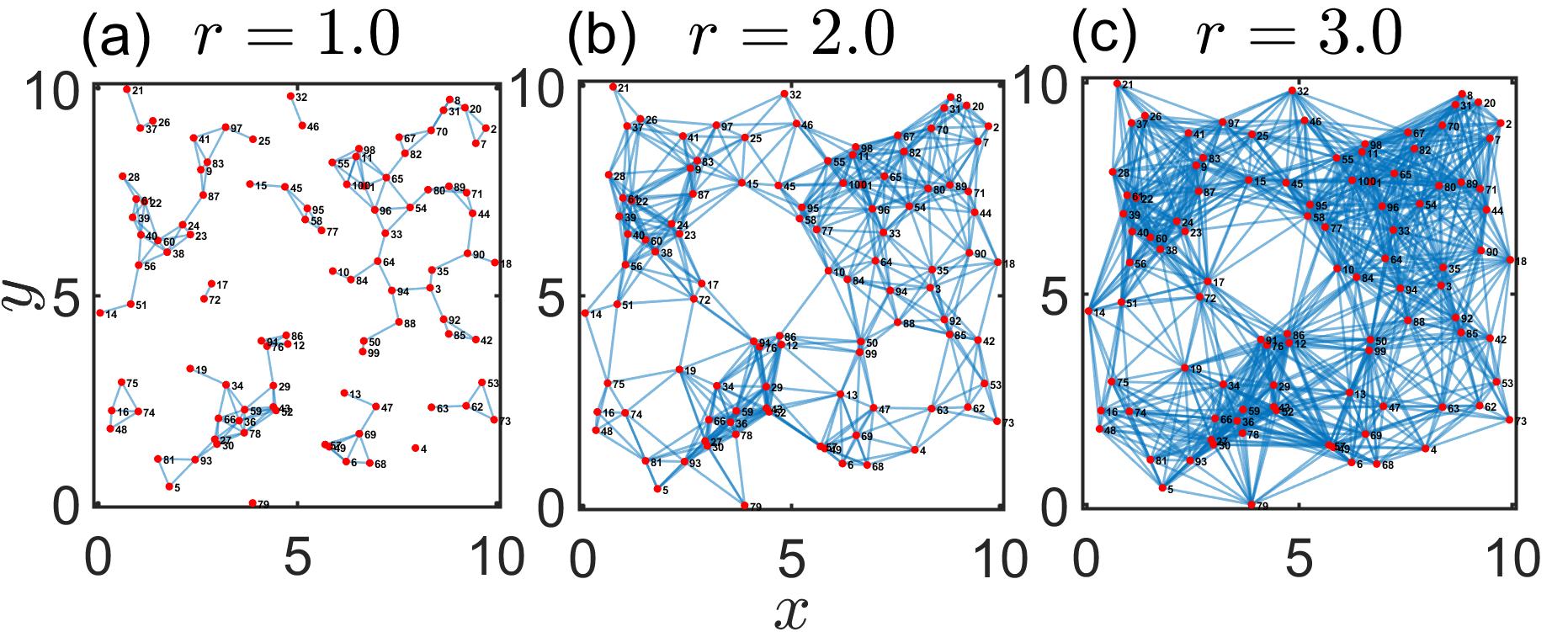}}
	\caption{Initial proximity graph $G_0$. The agents are distributed uniformly in $[0,10] \times [0,10]$ at initial time $t=0$. $G_0$ is demonstrated with the agents as vertices (red dots) and connection among them as edges (blue lines) for (a) $r = 1.0$, (b) $r = 2.0$, and (c) $r = 3.0$, respectively. The connectivity is measured for these graphs as (a) $\lambda_2=0$, (b) $\lambda_2=0.263$, and (c) $\lambda_2=1.78$. As $r$ increases, $\lambda_2$ also increases indicating that the graph becomes more strongly connected.}
	\label{fig2}
\end{figure}

\subsection{Cluster formations}
\begin{figure*}[ht]
	\centerline{
    \includegraphics[width = 2\columnwidth]{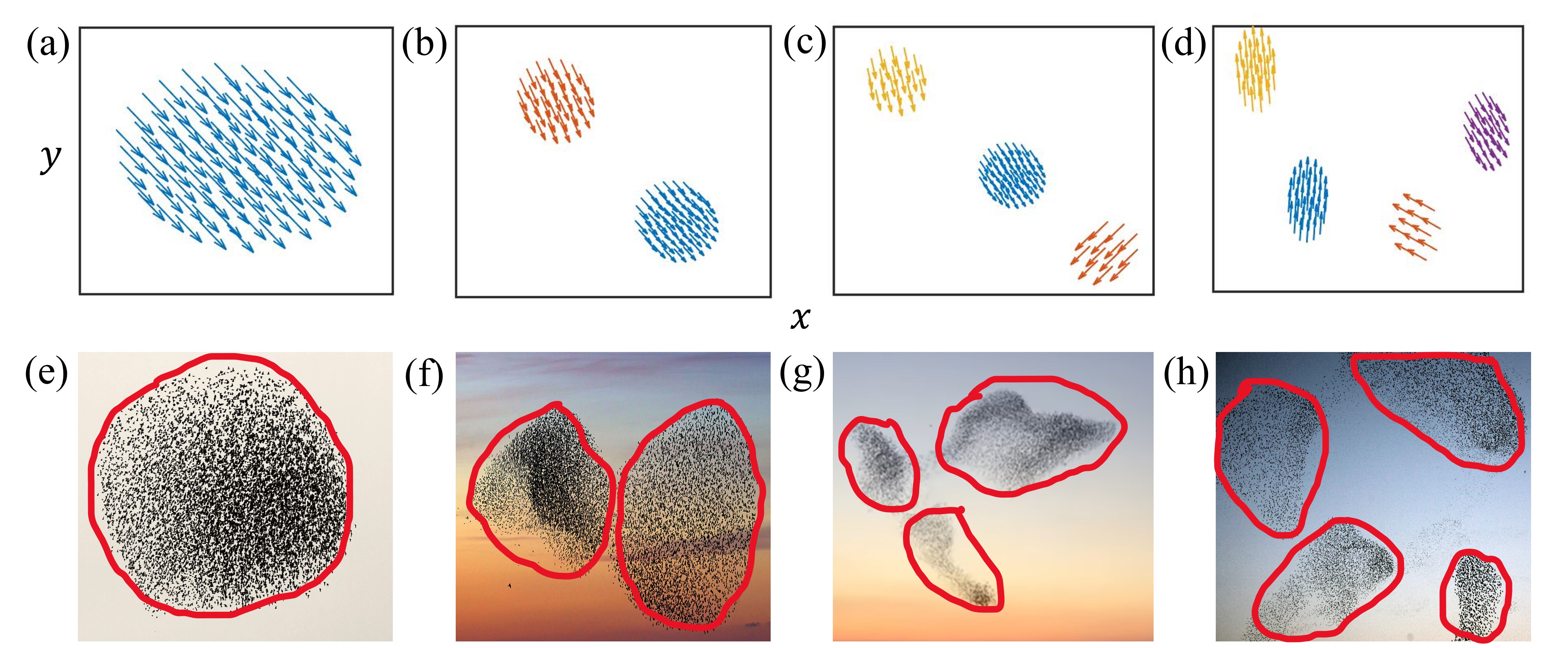}}
\caption{Formation of disjoint clusters depending on the interaction radius. Snapshots of the position of the agents at $t = 100$ time units are shown. (a) Four clusters for $r = 2.3$, (b) three clusters for $r = 2.6$, (c) two clusters for $r = 2.9$, and (d) single cluster for $r = 3.1$. Here, $v = 0.1$, and $K = 5.0$. The initial configurations are same in all four cases. Inside each cluster the agents are synchronized in directions, but desynchronized from cluster to cluster. Multimedia available online. (e)-(h) Instances of cluster formations in bird flocks seen in nature matching with our finding. Image source - Internet.}
\label{fig3}
\end{figure*}

A pivotal component of the model is the local interaction among the agents. Nearest neighbor interaction is the characteristic of many real world multi-agent systems where the entities affect and get influenced by only its local neighbors~\cite{reynolds1987flocks,okubo1986dynamical}. At time $t$, this neighbor set $N_i$ is determined by the interaction radius $r$. The neighbor set changes from time to time depending on the speed of the agent $v$. Let $G_0$ denote the proximity graph at $t=0$. If $r$ is small, then the initial proximity graph is disconnected, i.e., $\lambda_2(G_0)=0$ (see Fig.~\ref{fig2}(a)). This means there are at least two agents who are unable to influence each other directly (connected to each other) or through long-range interactions (connected through some other agents). But they are connected to their local neighbors, and as a result, they form clusters with them at large times. By clusters, we mean disconnected components of the proximity graph where there is no edge or link connecting them. Clearly, the clusters will reside at least $r$ distance away from one another. Beyond a critical value of $r$, the initial proximity graph becomes connected, i.e., $\lambda_2(G_0)>0$ (Fig.~\ref{fig2}(b)). In some previous studies on flocking, the connectivity of $G_0$ ensured the connectivity at all times~\cite{zhu2012flocking}. We observe this is not always true for our model. Even when $G_0$ is connected, the long time solution might not be connected. Look at Figs.~\ref{fig3}(a)-(c) (multimedia available online). In all the three cases, $\lambda_2(G_0)>0$, but they break in disjoint clusters at large time and become disconnected. The number of such disjoint clusters and number of agents inside each cluster depend crucially on the interaction radius $r$ and also on the initial position of the particles. When $r$ is reasonably large, the connectivity of $G_0$ is significantly strong (Fig.~\ref{fig2}(c)). Hence, they remain connected at all times (see Fig.~\ref{fig3}(d), multimedia available online). In this case, there is only one spatial cluster. These clusters are disc-shaped due to the choice of attraction and repulsion kernels in Eq.~(\ref{eq.12}). In Fig.~\ref{fig3}(e)-(h), we have shown real-world matches to the cluster formation of our model.

\par Now, we have the insight that depending on the value of $r$, the agents either form multiple disjoint disc-shaped clusters or remain connected inside a single cluster. But, the question remains whether the agents synchronize their direction so that the flocking situation is attained. Once the agents separate in disjoint clusters, they lose connection from other groups at all future instants of time. Since we are working in an unbounded domain, the clusters will keep moving away from each other with the mean directions of the respective clusters. Let $C$ denote the number of disjoint clusters that the agents get separated into. Also, let $M_k$, $m_k$, and $i_k$ denote the set of indices of the agents, number of agents, and the smallest index belonging to the $k$-th cluster, respectively, for $k=1,2,\ldots,C$. Trivially, the cardinality of the set $M_k$ is $m_k$, $i_k = \min M_k$, and $\sum_{k=1}^{C} m_k =N$. We define the averaged synchronization error as
\begin{equation}
    \theta_{err} = \frac{1}{C}\sum_{k=1}^{C} \Big \langle \frac{1}{m_k} \sum_{j \in M_k} |\theta_j - \theta_{i_k}| \Big \rangle,
    \label{eq.14}
\end{equation}
where $\langle \cdots \rangle$ stands for the time average. We say the directions are synchronized when $\theta_{err}$ is less than a threshold value $\theta_{T} = 10^{-5}$. This is calculated after discarding the transient time so that both the number of clusters ($C$) and the set of indices ($M_k$) remain constant for all $k$. It is to be noted that, $\theta_{err}=0$ does not necessarily mean all the agents' directions are same (except for when $C=1$). It implies that the agents inside each cluster are synchronized in their direction while they can be desynchronized from one cluster to another. In a nutshell, Eq.~(\ref{eq.14}) captures the local synchronization among agents rather than the global synchronization except for $C=1$. For $C=1$ and $\theta_{err}=0$ evidently signify that all the agents move in a single cluster and their directions are synchronized.

\par
\begin{figure}[t]
	\centerline{
    \includegraphics[width = \columnwidth]{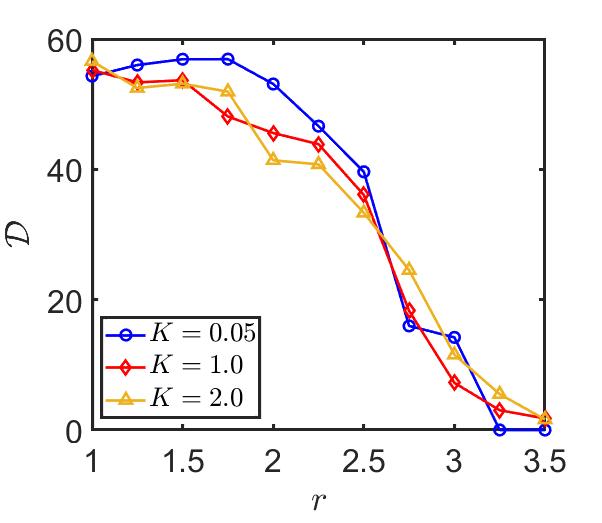}}
\caption{Maximum distance between clusters ($\mathcal{D}$) as a function of $r$. Blue circles, red diamonds, and yellow triangles stand for $K=0.05$, $K=1.0$, and $K=2.0$, respectively. Decreasing trend of $\mathcal{D}$ is observed with increasing $r$. With enhanced interaction radius the clusters lie closer to each other. $\mathcal{X}_k$s and $\mathcal{D}$ are calculated at $t=300$ time units. Each data point is achieved after taking the average over $10$ realizations of initial conditions. We use $v=0.1$ and $N=100$ here.}
\label{spread}
\end{figure}

 The interaction radius also controls the spread of the region that the agents occupy. Since they move in the unbounded plane, it is not guaranteed that the agents will be positioned close to each other at large times. We calculate the maximum distance between the center of positions of any two clusters. Let $\mathcal{X}_k$ denote the center of position of the $k$th cluster for $k=1,2,\ldots,C$ and $\mathcal{D}$ denote the maximum distance between them. Clearly, $\mathcal{D}=0$ when $C=1$. In Fig.~\ref{spread}, we plot the change in $\mathcal{D}$ as a function of $r$ for three different values of $K=0.05$ (blue circles), $K=1.0$ (red diamonds), and $K=2.0$ (yellow triangles). The decreasing nature of $\mathcal{D}$ with increasing $r$ for all $K$ values indicates that the clusters lie closer to each other when the value of $r$ grows.

\par As we can see that $r$ has a profound effect on the cluster formation, we try to find a lower bound on it so that $G_0$ is connected. For that, we assume that $N_s$ linearly spaced agents placed on a $2d$ square lattice containing $N_s$ lattice points where $N_s$ is the smallest square integer greater than or equal to $N$. Then it is easy to see that over such a formation the necessary condition for the proximity graph to be connected is $r>L/(\sqrt{N_s}-1)$. For $N=100$ with $L=10$, this becomes $r>1.1$. Even when the agents are uniformly distributed, this condition seems to work almost every time.

\subsection{Direction synchronization/flocking}
\begin{figure}[ht]
	\centerline{
		\includegraphics[width = \columnwidth]{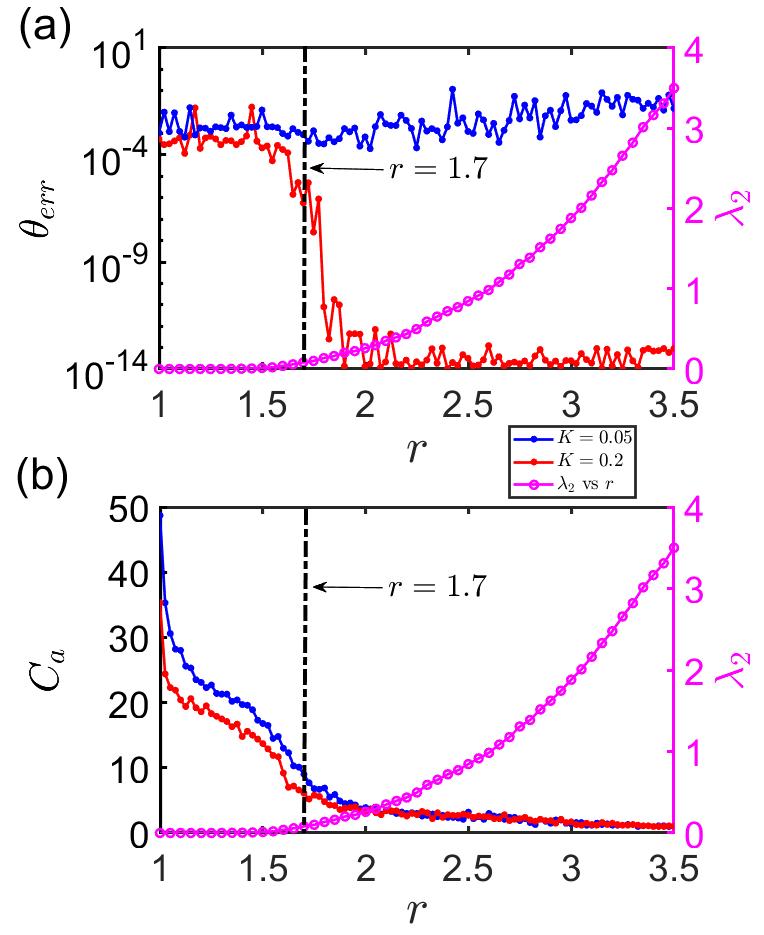}}
	\caption{$\theta_{err}$ and $C_a$ as functions of $r$. Simulations are performed for $K=0.05$ (blue dots) and $K=0.2$ (red dots). (a) Synchronization error $\theta_{err}$ shows distinct behaviors for the two $K$ values. The error is shown in the log scale here. (b) Average number of clusters $C_a$ shows quite similar trend. In the right $y$ axes of both (a) and (b), we plot the algebraic connectivity $\lambda_2(G_0)$ (pink dots). We use, $v=0.1$ here. Simulations are done for $t=500$ time units with $dt=0.01$. $\theta_{err}$ is time averaged over last $5\%$ data which is further averaged with $20$ different initial conditions. $C_a$ and $\lambda_2(G_0)$ are also averaged with $20$ trials each.}
	\label{fig4}
\end{figure}

So far, we have discussed the role of vision radius on the formation of spatial clusters. It still remains to be found out how the agents inside the clusters synchronize their directions. This is where the role of $K$ comes in. It minimizes the difference in directions between two agents as long as they are connected. The larger the value of $K$, the greater is the synchronization tendency. We have demonstrated the effect of $K$ in Fig.~\ref{fig4}(a) where we plot the synchronization error $\theta_{err}$ as a function of the interaction radius $r$ for two different values of $K$. The left and right $y$ axes stand for $\theta_{err}$ and $\lambda_2(G_0)$, respectively. The blue dotted curve is drawn for $K=0.05$ and it is seen that $\theta_{err}$ is always greater than $\theta_{T}$, which shows the directions never get synchronized. So, with $K=0.05$, flocking is never achieved even when $r$ is significantly large. The red dotted curve is drawn with a larger value of $K=0.2$. Now, we see that there is a significant drop of $\theta_{err}$ around $r \approx 1.7$. Beyond this value, $\theta_{err} < \theta_T$ and flocking is achieved. We denote this critical value as $r_c=1.7$. What actually happens around this $r_c$? Numerics reveal that the algebraic connectivity of the initial proximity graph ($\lambda_2(G_0)$) is approximately $0.1$ here. For $r>r_c$, $\lambda_2(G_0) \gtrapprox 0.1$ and the proximity graph is strongly connected to achieve flocking. In Fig.~\ref{fig4}(a), we have drawn $\lambda_2(G_0)$ as a function of $r$ with a pink dotted curve and the black dashed line indicates $r_c$.

\par Next, we calculate the average number of clusters $C_a$ while varying $r$. We take $20$ realizations of the initial conditions and average it to find $C_a$ in Fig.~\ref{fig4}(b). Here we see that both the curves for $K=0.05$ (blue dotted line) and $K=0.2$ (red dotted line) exhibit similar trends. Initially, $C_a$ is large as the interaction radius is small. It drops significantly around $r \approx r_c$ where the connectivity of the graph gets strong. $C_a$ tends to $1$ as $r \gg r_c$. The similar behaviors of the two curves in Fig.~\ref{fig4}(b) dictate that cluster formation is mainly determined by $r$, and $K$ does not play a significant role here, whereas, we have already found out that direction synchronization or flocking is influenced by both these parameters.

\par
\begin{figure}[t]
	\centerline{
		\includegraphics[width = \columnwidth]{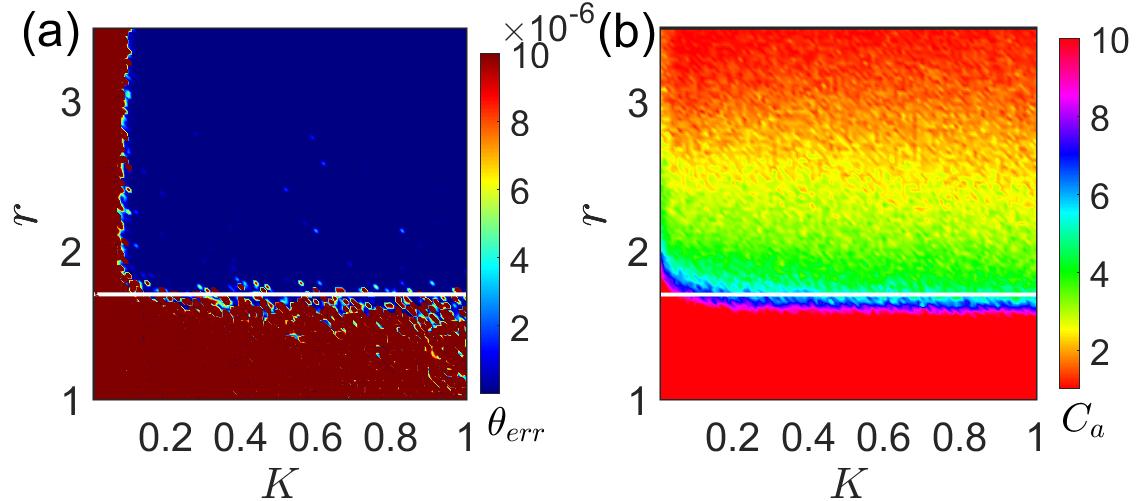}}
	\caption{Demonstration on $\theta_{err}$ and $C_a$ in the $K$-$r$ plane. We divide the region in $100\times 100$ mesh points. At each point we calculate the values of $\theta_{err}$ and $C_a$. (a) The $\theta_{err}$ values are delineated with the colorbar highlighting its range. (b) Variation of $C_a$ is depicted with its corresponding range in the colorbar. Simulation parameters: $t=500$, $dt=0.01$. Here, $v=0.1$, and $N=100$. Each data point is averaged over $10$ realisations of initial conditions.}
	\label{fig5}
\end{figure}

 To simultaneously study the effect of $K$ and $r$, we illustrate the behaviors of $\theta_{err}$ and $C_a$ in the $K$-$r$ parameter plane. Figure~\ref{fig5}(a) depicts the synchronization error in this plane with the maroon and blue colors representing the desynchronized ($\theta_{err} \ge \theta_T$) and synchronized ($\theta_{err} < \theta_T$) regions, respectively. For small values of $K$ ($<0.1$), the directions never get synchronized even with large $r$. Beyond $K=0.1$, synchronization starts to occur when $r>r_c$. The white solid line stands for $r=r_c$. In Fig.~\ref{fig5}(b), the average number of clusters is shown. Above the $r=r_c$ line, $C_a$ drops from a larger value to a value around $C_a \approx 4$. The effect of $K$ is minimal here, i.e., $C_a$ is almost independent of $K$.

\subsection{Effect of $v$}
\begin{figure}[ht]
	\centerline{
		\includegraphics[width = \columnwidth]{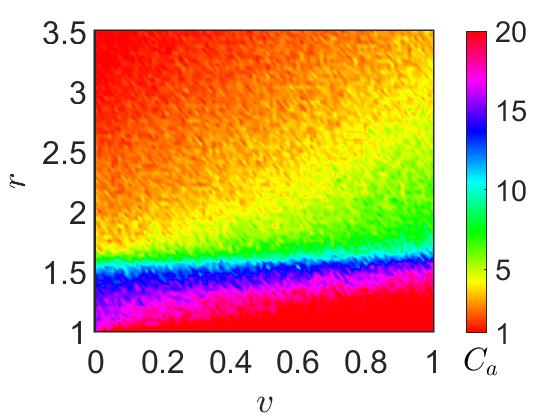}}
	\caption{Effect of $v$ on $C_a$. We vary $v$ and $r$ over suitable ranges and show the variation of $C_a$. Keeping $r$ fixed, if we move horizontally with increasing $v$, we find that $C_a$ also increases. For simulations we use $K=1.0$. Each data point is an average of $10$ realizations of initial conditions.}
	\label{fig6}
\end{figure}

The self-propulsion speed of the agents has a bearing on the ultimate fate of their swarming structure. Since the agents move in the unbounded plane, they can move away from each other very quickly when their directions are not synchronized, given $v$ is large. An increased value of $v$ means decrement of interaction time between the agents which eventually leads to a weaker coupling both in spatial position and direction. As we can predict, it is found that the number of synchronized clusters $C_a$ increases when we increase $v$. We delineate this scenario in Fig.~\ref{fig6}. The variation of $C_a$ is demonstrated in the $v$-$r$ parameter plane. It can be seen that, for a fixed value of the interaction radius, the average number of clusters increases when we increase the speed of the agents.

%%%%%%%%%%%%%%%%%%%%%%%%%%%%%%%%%%%%%%%%%%%%%%
\subsection{Effect of noise}

\begin{figure}[ht]
	\centerline{
		\includegraphics[width = \columnwidth]{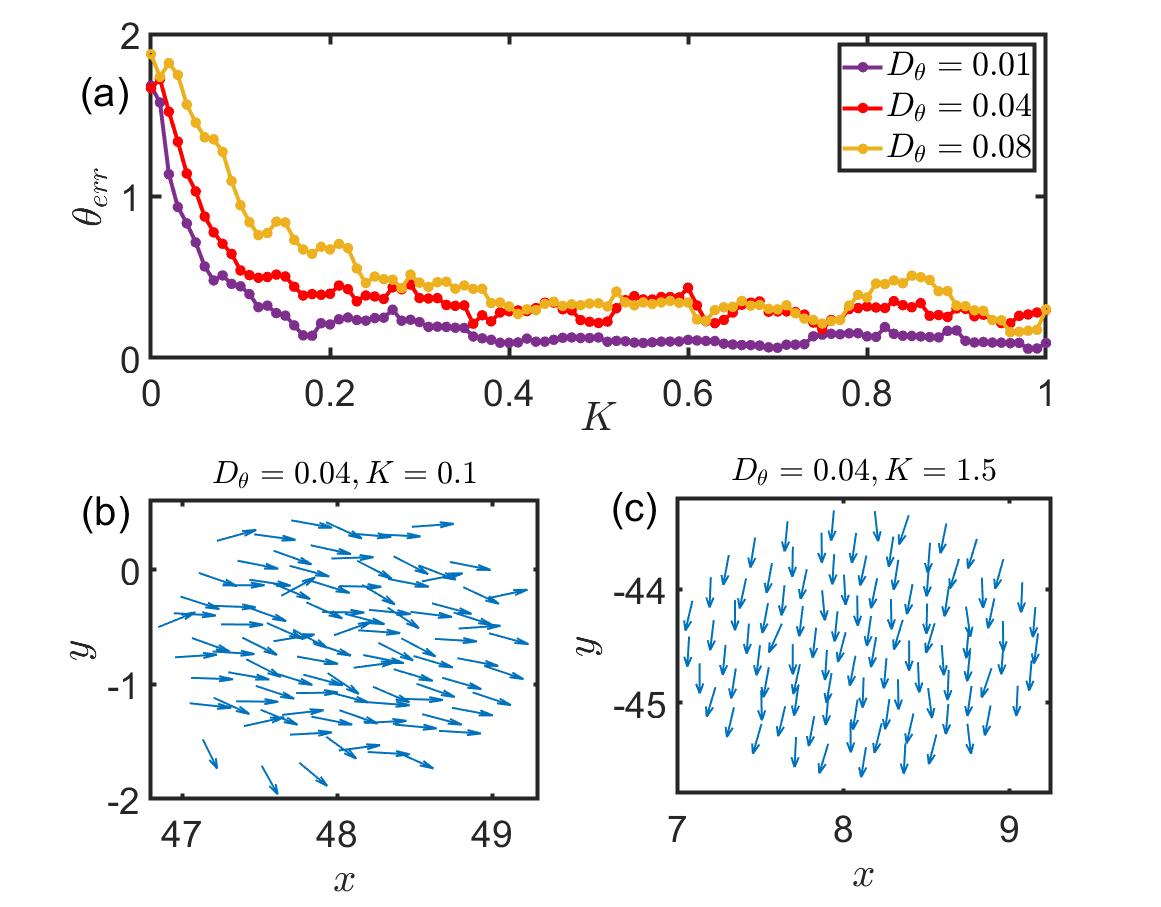}}
	\caption{Effect of noise on the directions of the agents. (a) $\theta_{err}$ is calculated as a function of $K$ for three different noise strengths $D_{\theta} = 0.01$ (purple), $0.04$ (red), and $0.08$ (yellow). Simulation parameters: $t=500$, $dt=0.01$. Here, we use $v=0.1$, $N=100$, and $r=5.0$. Each data point is an average of $20$ realizations of initial conditions. Snapshots of the agents for (b) $K=0.1$ when the directions are not synchronized, and (c) $K=1.5$ when the directions are synchronized. $D_{\theta}=0.04$ in both the cases. }
	\label{fig8}
\end{figure}

Noise is a key component of any real-world systems. It is significant as it captures the randomness, uncertainty, and variability of a system. Here, we introduce the noise term in the directional dynamics Eq.~\eqref{eq.13} of our model as
\begin{align}
	\dot{\theta}_{i} &=\frac{K}{|N_{i}|} \sum_{\substack{j = 1}}^{N} a_{ij} \sin(\theta_j - \theta_i) + \eta_i(t),	
	\label{eq.noise}
\end{align}
where $\eta(t)$ is white noise variable with zero mean and strength $D_{\theta}$ characterized by $\langle \eta_i(t) \eta_j(t')\rangle = 2D_{\theta} \delta_{ij} \delta(t-t')$. In presence of noise, the directions of the agents remain disordered and desynchronized. A larger coupling strength $K$ is required than the noiseless scenario to overcome the disordering effect of noise and achieve the flocking situation where the directions are synchronized. We use $D_{\theta}=0.01,0.04$, and $0.08$ and calculate the synchronization error $\theta_{err}$ as function of $K$ in each case. The result is presented in Fig.~\ref{fig8}(a). There is an overall decreasing trend of $\theta_{err}$ with increasing $K$ barring small fluctuations in between. The purple, red, and yellow curves are for $D_{\theta}=0.01$,$0.04$, and $0.08$, respectively. It is also seen that for a fixed value of $K$, a larger $D_{\theta}$ value yields a larger value of $\theta_{err}$ in most of the cases. In contrast to the noiseless scenario, here $\theta_{err} \approx 10^{-3}$ even with large coupling strengths and never drops below it due to the presence of uncorrelated noise. In Figs.~\ref{fig8}(b)-(c), snapshots of the agents are shown for a small coupling $K=0.1$ where the directions are disordered and for a large coupling $K=1.5$ where the directions are synchronized.

\color{black}
%%%%%%%%%%%%%%%%%%%%%%%%%%%%%%%%%%%%%%%%%%%%%%
\section{Conclusion}
Coordinated movements of animals are often found in nature. In this article, we have introduced a simple model which can capture their collective behaviors, mainly the two features, swarming and flocking. The agents interact with their local neighbors both spatially and in direction. They are assumed to move with a constant uniform velocity. We have relaxed the boundedness of the region of locomotion so that the agents can travel in the unbounded plane. The initial connectivity of the proximity graph impacts the long time behavior of the system. The connectivity strongly depends on the value of the interaction radius. For a small value of it, the proximity graph remains disconnected and the agents move on their own where their motion is not affected by others. When the interaction radius increases, the agents start to interact with each other and move in unison. For intermediate values of the interaction radius, we observe the agents break into spatially disjoint clusters. The number of such clusters and number of agents inside the clusters strongly depend on the value of the interaction radius as well as the choice of initial condition. Eventually all the agents move in a single cluster when the the interaction radius is very large. On the other hand, when we study the directional synchronization among the agents for flocking, we find that the directions get synchronized when the value of the directional coupling strength is greater than a critical threshold. The directional coupling strength does not affect the spatial cluster formations but it strongly impacts the synchronization of moving angles.

\par Flocking phenomena with consensus protocols have been widely studied in control theory while studying the coordinated behavior of mobile agents~\cite{olfati2007consensus,olfati2004consensus,jadbabaie2003coordination,zavlanos2007flocking}. A popular model for the evolution of flock is the Cucker-Smale model~\cite{cucker2007emergent} where velocity of the agents are time-varying and flocking is defined as the scenario of achieving a common velocity. The initial condition has been found to have a significant effect on the final outcome in many such studies. The connectivity of the proximity graph has also been in focus of most of these investigations. This has also been observed in our work. Our flocking problem has been improved by the addition of swarming dynamics in the spatial component. As a result, we observed the grouping phenomena. Different types of spatial interactions among the agents can give rise to other fascinating spatial structures which are commonly observed in the coordinated movements of birds and animals. To make the model more realistic, the effect of surrounding noise on the spatial components can be considered. We have briefly discussed the effect of noise on the directional dynamics and the results show that synchronization in directional variable is hindered by uncorrelated noise. Future research can also be carried out dealing with the analytical properties of the model for other types of noise.

%%%%%%%%%%%%%%%%%%%%%%%%%%%%%%%%%%%%%%%%%%%%%%
%\section*{\label{sec:level6}Supplementary Movies}
%We have made Movies for different cluster formations (snapshots are shown in Fig.~\ref{fig3}). Movie 1 shows emergence of four clusters for $r=2.3$. Movie 2 is for three clusters with $r=2.6$. Two cluster is shown in Movie 3 for $r=2.9$. In Movie 4, a single cluster formation can be seen where $r=3.1$. The parameter values $K=5.0$, $v=0.1$, and $N=100$ have been used in all four movies.

\section*{\label{sec:level7}Data Availability}
The data that support the findings of this study are available
from the corresponding author upon reasonable request.
%%%%%%%%%%%%%%%%%%%%%%%%%%%%%%%%%%%%%%%%%%%%%%
%\section*{Acknowledgements} 
%\section*{References}
%\bibliographystyle{apsrev4-1} % Tell bibtex which bibliography style to use
%\bibliography{flock}
%

\end{document}